\documentclass[11pt]{article}
\usepackage{enumitem} 
\usepackage[utf8]{inputenc}
\usepackage[T1]{fontenc}
\usepackage{lmodern}
\usepackage[margin=1in]{geometry}
\usepackage{amsmath}
\usepackage{amssymb}
\usepackage{booktabs}
\usepackage{tabularx}
\usepackage{ragged2e}
\usepackage{microtype}
\usepackage[super,numbers,sort&compress]{natbib}
\usepackage{authblk}

\usepackage{xcolor}
\usepackage{graphicx}
\usepackage{caption}
\usepackage{titlesec}
\titleformat{\subsection}[runin]{\normalfont\itshape}{\thesubsection}{0.5em}{}
\usepackage{hyperref}
\hypersetup{
  colorlinks=true,
  citecolor={blue!55!black},
  linkcolor={blue!55!black},
  urlcolor={blue!55!black}
}

\title{Do decisions about outliers and influential effects matter?\\[0.45em]
\Large Evidence from 358 behavioral science meta-analyses}

\author[a,b,c]{Tomas Havranek\thanks{Corresponding author: Tomas Havranek, Institute of Economic Studies, Faculty of Social Sciences, Charles University, Opletalova 26, 110 00 Prague, Czech Republic; \href{mailto:tomas.havranek@fsv.cuni.cz}{tomas.havranek@fsv.cuni.cz}. Pre-registration: \url{https://doi.org/10.17605/OSF.IO/97CMV}. Online appendix: \url{https://meta-analysis.cz/outliers}. Replication package: \url{https://doi.org/10.5281/zenodo.21216506}.}}
\author[a,c]{Zuzana Irsova}
\author[a]{Martina Luskova}
\author[d]{T. D. Stanley}

\affil[a]{Charles University, Prague}
\affil[b]{Centre for Economic Policy Research, London}
\affil[c]{Meta-Research Innovation Center at Stanford}
\affil[d]{Deakin University, Melbourne}

\date{}

\begin{document}
\maketitle

\begin{abstract}
\noindent Meta-analysts routinely face estimates that look too large or extreme. Yet, how to handle them is left to the reviewer's judgment. The methods for detecting such estimates are well known. What is missing is an informed assessment of how much alternative handling choices might change a meta-analysis' conclusions. We fill this gap by analyzing the effects of four pre-registered handling treatments across 358 behavioral science meta-analyses with at least ten estimates. Each outlier handling treatment is estimated by two estimators (random effects and unrestricted weighted least squares), and compared to the `do-nothing' baseline on three outcomes: the pooled effect, statistical significance, and whether the effect reaches the smallest effect size of interest ($|d| \geq 0.20$). Our entire analysis and comparison pipelines were pre-registered. Alternative outlier handling treatments have little effect on the meta-analysis mean as the median absolute change in Cohen's~$d$ is at most 0.047 and often much less. Yet, at least one of these four treatments in combination with one of these estimators reverses the statistical significance of 11.5\% of meta-analyses and the smallest-effect-of-interest assessment in 15.9\%. Winsorizing has the least effect and DFBETAS the most. Categorical changes are found almost entirely among results already close to the decision boundary; strongly significant results essentially never change. These findings give applied meta-analysts, methods specialists, and reviewers a reference point for how much this under-reported choice matters and provide yet another reason for meta-analysts to publicly pre-specify their methods and handling treatments.

\end{abstract}



\clearpage
\section{Introduction}
\label{sec:intro}

Nearly every meta-analyst faces an estimate that looks too extreme to leave alone. What to do about apparent outliers and highly influential estimates is one decision, among many, that meta-analysts must make. Researcher degrees of freedom are the source of well-documented bias and heterogeneity in primary studies.\cite{simmons2011,gelmanLoken2014} They do not disappear when results are pooled,\cite{olssonCollentine2025} and research synthesis adds yet another layer of choices that can affect summary results. A substantial literature maps how far results move across: specifications of a single dataset\cite{steegen2016,simonsohn2020,patel2015}, independent analysis teams\cite{silberzahn2018}, and decisions that are individually defensible but not interchangeable\cite{delgiudice2021}. Decisions about the handling of outlying, highly influential, or extreme estimates must be made in nearly every synthesis; yet, their handling treatments are seldom reported explicitly. These decisions and their effects on meta-analysis results are the focus of our study.

Methods for detecting `outliers' are well known. For the convenience of exposition, we include extreme and highly influential estimates as `outliers' in our discussions below, while fully recognizing and accommodating the differences between traditional outliers and influential data. Viechtbauer and Cheung\cite{viechtbauerCheung2010} recommended the calculation of studentized deleted residuals and DFBETAS in the random-effects model but were agnostic about what decisions meta-analysts should make after knowing the values of these diagnostic statistics. Viechtbauer and Cheung\cite{viechtbauerCheung2010} did ``not advocat(e) the routine deletion of outliers or influential studies''; rather, these diagnostic statistics should be ``used as part of sensitivity analyses'' (p.~122; parentheses added). Deletion of `outliers' is not to be routine. However, sensitivity analyses do not avoid the necessity that a decision must be made at some point by the meta-analyst about what to report. Detection may identify which estimates are unusual, but how to handle them (keep, drop, or resolve) is left to the reviewer's judgment. The Cochrane tutorial on sensitivity analysis explicitly recognizes that researchers may observe an ``obvious outlier'' and ``remove it to see its impact'', while acknowledging that what follows ``relies on judgment of researchers rather than objective assessment of statistics''\cite{aung2026}. The methods literature concedes that this handling choice can be consequential: ``different outlier detection methods could lead to fairly different conclusions''\cite{meng2024}. What has been missing is an assessment of the effects of `outlier' handling, empirically, at scale. That is, how much does outlier handling matter in practice across numerous meta-analyses?

The nearest meta-research assessment concerns a very different decision: which heterogeneity-variance estimator to use?\cite{langan2015} Langan, Higgins, and Simmonds\cite{langan2015} compared five heterogeneity-variance estimators across 12{,}894 Cochrane meta-analyses and found conclusions ``naively based on statistical significance (at a 5\% level)'' to be ``discordant for at least one pair of estimators in 7.5\% of meta-analyses''. Kontopantelis, Springate, and Reeves\cite{kontopantelis2013} re-analyzed an even larger Cochrane collection, 57{,}397 meta-analyses, using a bootstrapped DerSimonian--Laird approach. Where heterogeneity had been detected but ignored, 17\% to 20\% of the statistical conclusions changed; otherwise, the rate was 1\% to 3\%. The lesson of this prior meta-research is that pooled estimates typically move little; yet, conclusions can change far more frequently than what the reported type I error or confidence level would lead one to believe. We pose the same basic questions for the outlier and influence handling decisions, with one important difference. Decisions about heterogeneity estimators cause the reweighting of each estimate while keeping every estimate; decisions about `outliers' can change which estimates are analyzed. We also add a consideration about whether the pooled effect reaches a smallest effect size of interest (SESOI) in addition to whether it is statistically significant. Robust estimation that accommodates outlying studies is, of course, another option\cite{wang2025tmeta,noma2024}, but how often alternative approaches change conclusions in practice has not been evaluated.

To our knowledge, this is the first empirical assessment of outlier sensitivity on a large scale. Our approach is agnostic by design. We recommend no single handling treatment, but rather compare alternative handling choices relative to the `do-nothing' baseline of keeping all estimates in the meta-analysis. Our pre-registered treatments have all been used by us in prior published meta-analyses, but we do not agree on which to prefer. One of us reaches first for winsorizing, another for influence diagnostics such as DFBETAS. Thus, the rationale of this study was born. Across 358 behavioral science meta-analyses from psychology, psychotherapy, and exercise with at least ten estimates ($k \geq 10$), we pre-registered five handling treatments and stated the criteria used to assess each in advance\cite{pap2026}:

\begin{enumerate}[topsep=4pt,itemsep=1pt]
\item do nothing (the baseline);
\item drop the estimate with the most extreme value $|\text{Cohen's } d|$;
\item delete estimates whose studentized deleted residual exceeds 3 in absolute value;
\item winsorize at the 5th and 95th percentiles;
\item delete estimates whose $|\mathrm{DFBETAS}|$ exceeds $2/\sqrt{k}$.
\end{enumerate}

Each treatment is calculated by two estimators: REML random effects with Hartung--Knapp adjustment, and unrestricted weighted least squares (UWLS). Three outcomes are compared against the do-nothing baseline: the pooled effect size, its statistical significance, and whether it reaches a SESOI of $|d| \geq 0.20$. Because each meta-analysis is estimated under both estimators, 715 results are estimable under all five treatments. Unlike a multiverse analysis, which crosses many decisions to map the dispersion of results, we vary one planned handling treatment at a time against the baseline of `do nothing'. Within each estimator and outcome, the input data, thresholds, and edge rules are fixed. Any difference between a treated result and the do-nothing baseline is induced by the handling rule. Because every treatment, threshold, and edge rule was fixed by the pre-analysis plan before any result was seen, the conclusion reversal rates (where a result changes from statistically significant to nonsignificant or from SESOI to not SESOI or vice versa) cannot be a product of the flexibility of our analysis.

From this large set of meta-analyses and treatments, we find that meta-analysis' mean effect is rather robust to the handling choice (the median absolute change in $d$ never exceeds 0.047); however, the reversal of interpretations can be notable. At least one of the four alternative treatments for handling `outliers' changes statistical significance (SS) in 7.7\% of the 715 meta-analysis-estimator combinations and the SESOI status in 10.3\%. In other words, these five treatments agree in 92.3\% and 89.7\% cases, respectively, and the reversals come almost entirely from results that were already near the decision boundary. When we count reversals by either estimator at the meta-analysis level (358), the unit Langan and colleagues use, the SS reversal rate is 11.5\% for at least one treatment, compared with the 7.5\% for the heterogeneity decision\cite{langan2015}. The remainder of the article describes the data (Section~\ref{sec:data}), the design (Section~\ref{sec:methods}), the confirmatory results (Section~\ref{sec:results}), and what the rates mean for practice (Sections~\ref{sec:discussion} and~\ref{sec:conclusion}).

\section{Data}
\label{sec:data}

Our sample is drawn from the BEAR~v2 collection\cite{vanzwet2026} of behavioral science meta-analyses. It covers three fields with all estimates expressed as Cohen's~$d$. Psychology meta-analyses come from Sladekova and colleagues\cite{sladekova2022}, where Fisher-$z$-transformed correlations are converted to $d$. Psychotherapy meta-analyses come from the Metapsy database\cite{plessen2023,metapsy}, where effects are reported as Hedges' $g$ or standardized mean differences; it is the only field containing a study identifier. The exercise meta-analyses come from Barto\v{s} and colleagues\cite{bartos2025exercise}, where effects are reported as standardized mean differences.

Our fixed dataset contains 18{,}170 estimates. We focus on meta-analyses with at least ten estimates ($k \geq 10$), below which influence diagnostics have too few points to be reliable. Altogether, there are 361 such meta-analyses (259 psychology, 20 psychotherapy, 82 exercise). Three of the exercise meta-analyses (Barto\v{s}/178, 180, and 224) cannot be calculated by random effects because the REML Fisher scoring algorithm does not converge. We drop them so that every meta-analysis that we report enters on the same footing. This leaves 358 meta-analyses (259 psychology, 20 psychotherapy, 79 exercise) containing 16{,}664 estimates. Table~\ref{tab:corpus} reports the per-field descriptives (the median $k$ and its quartiles, the median absolute meta-analysis mean, and the share of $I^2$ above 75\%). These descriptives motivate the rest of the analysis, showing that the three fields differ sharply in scale, effect size, and heterogeneity. Psychology and psychotherapy both tend to have medium effect sizes, in the $0.4$--$0.5$ $d$ range, but the psychotherapy meta-analyses are notably larger, in other words have more estimates per meta-analysis (median $k$ = 127 vs.\ 23). Exercise meta-analyses have both smaller average effects and fewer estimates. So, any one of these collections of meta-analyses alone would provide a thin basis for generalizations.

Two features of our fixed data are worth pointing out, because they may influence the analysis. First, thirty-seven psychology estimates from nine meta-analyses have very extreme values, $|d| > 10$, arising from the Fisher-$z$ conversion of near-unit correlations. All thirty-seven pass the registered data-validity filter, which drops any correlation at or above 0.999 in absolute value. The psychology source has no study identifiers, so we cannot trace the recorded correlations back to the primary studies. Drop-extreme, which selects on magnitude alone, is the one treatment certain to remove one of them, but doing so changes neither the SS nor the SESOI status in any of these nine. Second, 421 distinct combinations of $d$ \& $SE$ values occur more than once within the same meta-analysis, spread over 61 meta-analyses. Some of these combinations recur several times, so together they account for 664 rows beyond the first occurrence of each. Three of us compiled the exercise collection, and checking those sources shows that the matching rows are distinct subsample estimates whose reported values coincide after rounding. The psychology source has no study identifiers, so we could not verify the provenance of its matching values, and we did not trace the psychotherapy matches back to their primary reports. We analyze the data as compiled because removing the matching rows would discard genuine estimates (at least in the exercise meta-analyses) and would become a different dataset from the one that we registered and have frozen. We discuss this as a limitation in Section~\ref{sec:discussion}.

The pre-analysis plan, the fixed dataset, and exact data checksum were registered on OSF\cite{pap2026} before any results were seen (Section~\ref{sec:methods}). We interpret our findings as descriptive of behavioral science meta-analyses with $k \geq 10$. Section~\ref{sec:methods} sets out the five handling treatments (including `do nothing'), two estimators, and three outcomes.

\begin{table}[t]
\centering
\caption{Descriptive Statistics by Behavioral Science Field}
\label{tab:corpus}
\small
\begin{tabularx}{\textwidth}{@{}l r r r X r r@{}}
\toprule
Field & \shortstack[r]{Meta-\\analyses} & \shortstack[r]{Median $k$\\(p25, p75)} & \shortstack[r]{Estimates\\($\Sigma k$)} & Base metric & \shortstack[r]{Median\\$|d|$} & \shortstack[r]{$I^2 > 75\%$} \\
\midrule
Psychology     & 259 & 23 (14, 40)   & 10{,}729 & Fisher-$z$ correlations $\to d$ & 0.421 & 61.0\% \\
Exercise       &  79 & 15 (12, 20.5) &  1{,}540 & SMD (as $d$)                    & 0.285 & 39.2\% \\
Psychotherapy  &  20 & 127 (87, 279) &  4{,}395 & Hedges $g$ / SMD (as $d$)       & 0.494 & 50.0\% \\
\midrule
\textbf{Overall} & \textbf{358} & \textbf{21 (14, 40)} & \textbf{16{,}664} & mixed (see per-field) & \textbf{0.400} & \textbf{55.6\%} \\
\bottomrule
\end{tabularx}

\vspace{0.4em}
\parbox{\textwidth}{\footnotesize\itshape Notes: $k$ = number of estimates in a meta-analysis; fields are ordered by number of meta-analyses. $\Sigma k$ counts total estimates analyzed. Median $|d|$ is the median absolute pooled effect under random-effects (RE) with no outlier or influence handling (i.e., the `do-nothing' baseline). $I^2 > 75\%$ is the share of meta-analyses with high heterogeneity.}
\end{table}

\section{Methods}
\label{sec:methods}

For each meta-analysis, we apply four `active' pre-specified handling treatments (not including the `do nothing' baseline) using two estimators and on three outcomes with the `do nothing' baseline. For each comparison, the input data, thresholds, edge rules, estimator, and outcome definition are held fixed. Any change from baseline therefore reflects the marginal effect of that handling treatment.\looseness=-1

\subsection{Treatments}

Each of the five treatments is an `outlier' (again, including both extreme and influential estimates) treatment that a meta-analyst might apply to the effect-size estimates before pooling them to calculate the meta-analysis estimate of the mean effect. \textit{None} is the do-nothing baseline where no estimate is removed or altered, and the other four treatments are measured against it. \textit{Drop-extreme} removes the single estimate with the largest $|d|$; if two estimates tie on $|d|$, we remove the one with the larger standard error, and if they still tie, the earlier row. \textit{Drop-extreme} is the crudest treatment we consider because it ignores precision. We include it to bound how far a simple magnitude rule can move a result. \textit{Studentized} removes every estimate whose studentized deleted residual exceeds $3$ in absolute value, then recalculates each estimator. \textit{Winsorize} replaces effect values below the 5th percentile within a meta-analysis with the 5th-percentile value and those above the 95th with the 95th-percentile value (R's default type-7 quantile). Unlike the other treatments, \textit{Winsorize} changes but does not delete effects, leaving $k$ and SEs unchanged. \textit{DFBETAS} removes every estimate whose $|\mathrm{DFBETAS}|$ from the meta-analysis mean exceeds $2/\sqrt{k}$ in absolute value, then recalculates the meta-analysis estimate of the mean effect. DFBETAS is the change in the mean effect, in standard-error units, when that estimate is left out. For the random-effects model, the studentized and DFBETAS diagnostics are the leave-one-out quantities from the \texttt{metafor} package\cite{viechtbauer2010}, which then refit REML and re-estimate $\tau^2$. For the unrestricted weighted least squares, the studentized deleted residuals and DFBETAS of the no-intercept regression are calculated as described below.

\subsection{Estimators}

Each baseline and treatment is estimated under two conventional estimators. We use only two to focus on the treatments. \textit{Random effects (RE)} is the reference. RE estimates the between-study variance $\tau^2$ by restricted maximum likelihood with the Hartung--Knapp--Sidik--Jonkman (HKSJ) adjustment. HKSJ assumes a $t$ distribution with one fewer degree of freedom than the number of estimates fitted.\cite{hartung2001,sidik2002,inthout2014} We use HKSJ as it is widely considered to produce better standard errors than other conventional RE standard errors. REML plus Hartung--Knapp usually gives wider, more accurate confidence intervals. \textit{Unrestricted weighted least squares (UWLS)} is the inverse-variance-weighted mean, $\widehat{\theta}_{\mathrm{UWLS}} = \sum_i d_i v_i^{-1} / \sum_i v_i^{-1}$, with a standard error inflated by the observed multiplicative heterogeneity. Equivalently, it is the slope of the no-intercept least-squares regression of $d_i/\mathrm{SE}_i$ on $1/\mathrm{SE}_i$, tested with a $t$ distribution with one fewer degree of freedom than the number of estimates fitted.\cite{stanley2015,stanley2023,stanley2026uwls,stanleyHavranek2025} As a least-squares estimator, UWLS has the smallest variance among all linear unbiased estimators as proved by the Gauss--Markov theorem without requiring normality.\cite{stanley2026uwls} It also weights and responds to heterogeneity differently from RE; comparing the two shows whether a handling treatment matters more under one estimator than the other. UWLS's residuals and DFBETAS used by treatments 3 and 5, respectively, are those of this simple OLS regression.\looseness=-1

\subsection{Outcomes}

For every combination of meta-analysis and estimator, we compare each `active' treatment (2--5) with the baseline under the same estimator, on three outcomes. The first outcome is the change in the meta-analysis mean, measured by the absolute size $|\Delta d|$ for $\Delta d = \widehat{\theta}_{\text{treated}} - \widehat{\theta}_{\text{baseline}}$. The second is statistical significance. Using a two-sided test at $\alpha = .05$ with $k_t - 1$ degrees of freedom, each cell falls into one of four states relative to baseline: significant before and after, non-significant before and after, becomes significant, or becomes non-significant. A $p$-value of exactly $.05$ is treated as non-significant. In practice, no such tie occurs. At full precision, none of our $p$-values equals $.05$; the closest is $.0501$. The third is whether the meta-analysis mean is as large as the smallest effect size of interest (\textit{SESOI}); that is, whether $|\widehat{\theta}| \geq c$ with $c = 0.20$ (i.e., Cohen's ``small effect''\cite{cohen1988}) as the main threshold and $c = 0.10$ as a sensitivity check. We call a change from either side a reversal or switch, and the rate we report is a switching rate relative to doing nothing. No diagnostic can identify which estimates are the ``true'' `outliers'. Viechtbauer and Cheung\cite{viechtbauerCheung2010} are explicit (p.~122) that these diagnostics are best used as a sensitivity analysis rather than to make routine deletions. A reversal therefore records that a meta-analytic conclusion under the specified estimator is sensitive to a specific and defensible handling choice. We cannot know whether any such changes improve the conclusion or not.

\subsection{`Outlier' identification}

The two cut-offs we use are widely accepted and are fixed in our pre-analysis plan. Our pre-registered studentized deleted residual cut-off of $3$ is stricter than the $1.96$ that Viechtbauer and Cheung\cite{viechtbauerCheung2010} used for illustration. We intentionally chose the larger value so that only truly unusual residuals are flagged. $|\mathrm{DFBETAS}|$ larger than $2/\sqrt{k}$ are flagged as this is widely accepted and suggested by Belsley, Kuh, and Welsch\cite{belsley1980}. $k$ is counted before the treatment is applied. This treatment is more sensitive for large $k$ than the fixed value of $1$ that Viechtbauer and Cheung\cite{viechtbauerCheung2010} suggest. Both criteria are applied to the absolute value, as specified in the pre-analysis plan. Thus, unusually negative and unusually positive estimates are flagged alike. There is no single correct threshold for any of these methods, and different detection methods can lead to different conclusions\cite{meng2024}. The absence of a single accepted standard is what makes comparing across defensible treatments worthwhile and why we vary these cut-offs in registered robustness checks rather than relying on one specific number.

\subsection{Pre-registration and robustness}

Our full study design was registered on the Open Science Framework\cite{pap2026} (\url{https://doi.org/10.17605/OSF.IO/97CMV}) before any outcomes were computed. Every treatment, estimator, outcome, threshold, and edge rule was fixed in advance; thus, these findings reflect no researcher degrees of freedom or forking paths.\cite{simmons2011,gelmanLoken2014} Statistics that combine the four treatments, such as the share of results changed by at least one treatment and the Fleiss $\kappa$, are descriptive summaries of these registered comparisons, not additional analyses. Five robustness variants were registered alongside the main analysis. The first restricts the sample to $k \geq 20$, where the residual and influence diagnostics are more reliable. The second lowers the floor to $k \geq 5$, to gauge the sensitivity of very small meta-analyses. The third recomputes treatments 3 and 5 at nearby cut-offs (studentized residual $> 2.5$; DFBETAS $> 3/\sqrt{k}$). The fourth collapses estimates from the same primary study to their inverse-variance mean before any `outlier' treatment to address potential dependency within studies. Then, $k$ = the number of studies in this meta-analysis. This robustness check applies only to psychotherapy because it alone has a study identifier. Lastly, the SESOI sensitivity threshold is changed to $c = 0.10$ and reported with the results (Table~A.9).

A cell is recorded as not estimable and left out rather than counted as a change if the treatment leaves fewer than three estimates or if the model returns no finite estimate or standard error. When REML returns a between-study variance of zero, we keep the registered Hartung--Knapp inference rather than switching to a fixed-effect interval. We investigate 358 distinct meta-analyses, not repeated analyses of a single dataset; thus, the dependency that many-analyst and multiverse designs must accommodate does not arise here.\cite{bartos2025multiverse} Some primary studies may still enter more than one meta-analysis, and only psychotherapy carries a study identifier, so we cannot rule that out.

\section{Results}\label{sec:results}

We report the three registered outcomes: how far the meta-analysis mean moves, how often its statistical significance changes, and how often the smallest-effect-of-interest verdict switches. Each treatment is measured against ``none'', the do-nothing baseline, so the results below are relative to leaving the estimates alone. Each of the 358 meta-analyses contributes up to two cells, one under random-effects (RE) and one under unrestricted weighted least squares (UWLS). Of the 716 possible combinations (i.e., cells), 715 are estimable under all five treatments. Summaries that pool the three fields are weighted by the composition of the sample, in which psychology contributes 259 of the 358 meta-analyses; the pre-registered breakdown by field is reported in the Online Appendix (Table~A.1).

\begin{table}[t]
\centering
\caption{Median Effects of Outlier Treatments}
\label{tab:effectshift}
\small
\setlength{\tabcolsep}{4pt}
\begin{tabularx}{\textwidth}{@{}X r r r r r r@{}}
\toprule
 & \multicolumn{2}{c}{Pooled $|\Delta d|$} & \multicolumn{2}{c}{RE $|\Delta d|$} & \multicolumn{2}{c}{UWLS $|\Delta d|$} \\
\cmidrule(lr){2-3}\cmidrule(lr){4-5}\cmidrule(lr){6-7}
Treatments & Median & (p25, p75) & Median & (p25, p75) & Median & (p25, p75) \\
\midrule
Drop-extreme                   & 0.024 & (0.011, 0.054) & 0.033 & (0.017, 0.064) & 0.016 & (0.007, 0.039) \\
$|$Studentized residual$| > 3$ & 0.006 & (0.000, 0.055) & 0.000 & (0.000, 0.051) & 0.010 & (0.000, 0.056) \\
Winsorize 5/95                 & 0.006 & (0.002, 0.015) & 0.008 & (0.003, 0.022) & 0.004 & (0.002, 0.011) \\
$|$DFBETAS$| > 2/\sqrt{k}$     & 0.045 & (0.017, 0.097) & 0.047 & (0.017, 0.106) & 0.043 & (0.018, 0.087) \\
\bottomrule
\end{tabularx}

\vspace{0.4em}
\parbox{\textwidth}{\footnotesize\itshape Notes: $|\Delta d| = |\text{pooled }d\text{ under alternative outlier treatments} - \text{pooled }d\text{ under do nothing}|$. The pooled columns aggregate over both estimators (RE and UWLS); the RE and UWLS columns report each estimator separately. Values are medians over estimable cells with dispersion summarized by the 25th and 75th percentiles.}
\end{table}

\subsection{How much the estimated mean effect moves}

Table~\ref{tab:effectshift} reports the change in the pooled effect. Each row is one of the four active treatments. The first three columns give the change $|\Delta d|$ pooled over both estimators, as the median and its 25th and 75th percentiles. The remaining columns report the same summaries separately for RE and UWLS. The point estimate barely moves. Pooled over both estimators, the median absolute change $|\Delta d|$ is 0.024 under drop-extreme, 0.006 under the studentized-residual treatment, 0.006 under winsorizing, and 0.045 under DFBETAS. This ordering is expected. Drop-extreme deletes the single largest estimate and winsorizing only cuts off the tails. Winsorizing has the least effect (interquartile range 0.002 to 0.015), and DFBETAS removes the most influential points (IQR $= 0.097 - 0.017 = 0.08$). The studentized deleted RE median is exactly 0 because more than half of RE cells contain no estimate with $|t| > 3$; therefore, studentized deleted residuals most often do nothing at all when using RE. For UWLS, studentized deleted residuals have a small but non-zero effect, median $=$ 0.010. How often each treatment acts, and how many estimates it removes or winsorizes, is tabulated in the Online Appendix (Table~A.2); the full distribution of the pooled-effect changes is plotted there (Figure~A.1). The deposited signed changes show that the treated and baseline means have opposite signs in 57 of the 2{,}862 estimable treatment-versus-baseline comparisons (2.0\%, spread over 18 meta-analyses). These are all crossings near zero; in no case are both means as large as 0.10 in absolute value.

Even the largest median shift, 0.045 under DFBETAS, is less than one-twentieth of a standardized mean difference or 11.3\% of the typical meta-analysis mean (median absolute value is 0.400). Typically, how one handles `outliers' has little practical effect on the magnitude of the estimated mean effect. Nonetheless, how this effect is interpreted (statistically significant or larger than SESOI) is more notably affected.

\begin{table}[t]
\centering
\caption{Changes in statistical significance and smallest effect size of interest}
\label{tab:verdicts}
\footnotesize
\setlength{\tabcolsep}{3.5pt}
\begin{tabularx}{\textwidth}{@{}X rrr rrr rrr@{}}
\toprule
 & \multicolumn{3}{c}{RE} & \multicolumn{3}{c}{UWLS} & \multicolumn{3}{c}{Pooled} \\
\cmidrule(lr){2-4}\cmidrule(lr){5-7}\cmidrule(lr){8-10}
\multicolumn{10}{@{}l}{\textbf{Panel A: statistical significance} (SS; two-sided $p < 0.05$)} \\
Treatment & $\to$sig & $\to$non & Either & $\to$sig & $\to$non & Either & $\to$sig & $\to$non & Either \\
\midrule
Drop-extreme                   & 6 & 8 & 3.91\% & 7 & 3 & 2.79\% & 13 & 11 & 3.35\% \\
$|$Studentized residual$| > 3$ & 6 & 4 & 2.80\% & 8 & 2 & 2.79\% & 14 &  6 & 2.80\% \\
Winsorize 5/95                 & 9 & 0 & 2.52\% & 8 & 0 & 2.23\% & 17 &  0 & 2.38\% \\
$|$DFBETAS$| > 2/\sqrt{k}$     & 9 & 7 & 4.47\% & 16 & 7 & 6.42\% & 25 & 14 & 5.45\% \\
\midrule
\multicolumn{10}{@{}l}{\textit{At least one treatment changes SS in 7.7\% of results (55 of 715); $\kappa = 0.87$.}} \\
\midrule
\multicolumn{10}{@{}l}{\textbf{Panel B: smallest effect size of interest} (SESOI; $|\text{pooled } d| \geq 0.20$)} \\
Treatment & $\to$reach & $\to$below & Either & $\to$reach & $\to$below & Either & $\to$reach & $\to$below & Either \\
\midrule
Drop-extreme                   & 1 & 23 & 6.70\% &  1 & 17 & 5.03\% &  2 & 40 & 5.87\% \\
$|$Studentized residual$| > 3$ & 2 & 13 & 4.20\% & 11 & 15 & 7.26\% & 13 & 28 & 5.73\% \\
Winsorize 5/95                 & 1 &  5 & 1.68\% &  4 &  3 & 1.96\% &  5 &  8 & 1.82\% \\
$|$DFBETAS$| > 2/\sqrt{k}$     & 4 & 20 & 6.70\% & 17 & 10 & 7.54\% & 21 & 30 & 7.12\% \\
\midrule
\multicolumn{10}{@{}l}{\textit{At least one treatment changes SESOI in 10.3\% of results (74 of 715); $\kappa = 0.87$.}} \\
\bottomrule
\end{tabularx}

\vspace{0.4em}
\parbox{\textwidth}{\footnotesize\itshape Notes: Entries are changes relative to ``none'', the do-nothing baseline. ``Either'' is the sum of the two changes as a percent of estimable cells (also called ``discordance''). Treatments are the same as in Table~\ref{tab:effectshift}. Panel A. Statistical significance (SS) is evaluated by two-sided $p < 0.05$. $\to$sig counts results that become SS; $\to$non counts those that become non-SS. Panel B: the smallest effect size of interest (SESOI) is reached iff $|\text{pooled } d| \geq 0.20$. $\to$reach counts results that become SESOI; $\to$below the reverse.}
\end{table}

\subsection{Whether the mean effect is statistically significant}

Panel A of Table~\ref{tab:verdicts} reports the change in statistical significance (SS). Over the 715 combinations of estimators and meta-analyses, at least one of the four active treatments changes the two-sided $p < 0.05$ finding in 7.7\% of the estimable cases (55 of 715). The five treatments agree in the remaining 92.3\% (Fleiss $\kappa = 0.87$). Pooled over both estimators, the share of results that change SS status is 3.35\% under drop-extreme, 2.80\% under studentized deleted, 2.38\% under winsorizing, and 5.45\% under DFBETAS. The at-least-one rate is computed over the 715 cells estimable under all five treatments; each treatment's own rate is computed over all cells estimable under that treatment. DFBETAS changes SS most often; winsorizing the least, reflecting the same relative sensitivity as seen by the change in mean effect. Winsorizing only goes one way, toward significance.

The change of SS varies by estimator. Some treatment switches RE's SS in 6.44\% of its results and UWLS in 8.94\%. Thus, UWLS is the more sensitive. Nor is switching directionally symmetric. Of the 55 changed cells, 36 move from non-significant to significant. When extreme estimates are dropped or downweighted, both noise and signal decrease with ambiguous net results on SS. 11.5\% (41 of 358) of these meta-analyses have either RE or UWLS changing SS under some treatment. Langan and colleagues made similar meta-analysis comparisons and found significance-based conclusions discordant across heterogeneity estimators in 7.5\% of their cases.\cite{langan2015}

\subsection{Larger than the smallest effect size of interest (SESOI)}

Panel B of Table~\ref{tab:verdicts} reports whether the estimated mean effect is at least as large as the smallest effect size of interest, $|d| \geq 0.20$. This is a distinct question from significance, and its verdict reverses somewhat more often. At least one treatment changes the SESOI status in 10.3\% of the 715 meta-analysis-estimator combinations (74 of 715), but all five agree in the remaining 89.7\% (Fleiss $\kappa = 0.87$). That is, roughly one result in ten is size-sensitive.

SESOI's prevailing direction is the `opposite' of SS. Of the 74 changed cells, 50 fall below 0.20 under some treatment and only 24 are pushed above it. In this sample, outlier treatments more often shrink effects than enlarge them. Taken alone, RE changes the SESOI status in 8.96\% of its results under some treatment and UWLS in 11.73\%. Again, UWLS is more sensitive than RE. At the meta-analysis level, which counts a change under either estimator, 15.9\% of meta-analyses change SESOI status (57 of 358) by some `outlier' treatment. For the registered 0.10 sensitivity threshold, the corresponding rate among the 715 combinations falls from 10.3\% to 6.4\% (Online Appendix, Table~A.9). Further results with all three outcomes broken down by field are reported in the Online Appendix (Table~A.1); the full significance and SESOI transition counts are plotted there (Figure~A.2).

\begin{table}[t]
\centering
\caption{Registered robustness findings}
\label{tab:robustness}
\small
\begin{tabularx}{\textwidth}{@{}X r r r r r@{}}
\toprule
Specification & \shortstack[r]{Meta-\\analyses} & Cells & \shortstack[r]{$\geq 1$ treatment\\changes SS} & \shortstack[r]{$\geq 1$ treatment\\changes SESOI} & \shortstack[r]{Not\\estimable} \\
\midrule
Main (all $k \geq 10$)             & 358 &  715 &  7.69\% & 10.35\% & 2 \\
$k \geq 20$                        & 190 &  379 &  5.01\% &  7.65\% & 2 \\
$k \geq 5$                         & 521 & 1{,}040 & 13.46\% & 13.46\% & 3 \\
Alternative cutoffs                & 358 &  715 &  6.85\% & 10.63\% & 3 \\
One estimate per study$^{\dagger}$ & 358 &  715 &  7.69\% & 10.21\% & 2 \\
\bottomrule
\end{tabularx}

\vspace{0.4em}
\parbox{\textwidth}{\footnotesize\itshape Notes: Each row reports a pre-registered variation to our ``Main'' analyses. ``Cells'' are estimable combinations of estimators and meta-analyses. ``$\geq 1$ treatment changes SS'' is the share of cells where at least one of the four active treatments changes SS relative to none; ``$\geq 1$ treatment changes SESOI'' is the corresponding share for the $|d| \geq 0.20$. $^{\dagger}$The one-estimate-per-study variation applies only to the 20 psychotherapy meta-analyses, because only it has a study identifier; the other fields are not affected. Estimates from the same study are first collapsed to their inverse-variance mean, and $k$ then counts studies.}
\end{table}

\subsection{Robustness}

We pre-registered five robustness variants. Table~\ref{tab:robustness} reports the four that change the sample or the cut-offs, one per row, below the main analysis. The fifth lowers the SESOI threshold to 0.10 and is reported above with the SESOI results (Table~A.9). The reversal rates increase with smaller meta-analysis sizes ($k$). Restricting to $k \geq 20$ with 190 meta-analyses lowers rates to 5.01\% (SS) and 7.65\% (SESOI), while adding smaller meta-analyses, $k \geq 5$, raises both to 13.46\%. That the two rates are identical is a coincidence; both are 140 of 1{,}040 cells, and the two sets of changed cells overlap in only 49 cases. This pattern is to be expected as outlier handling will have more leverage to affect meta-analysis findings when there are fewer estimates (smaller $k$). The alternative cutoffs (a studentized threshold of 2.5 and a DFBETAS threshold of $3/\sqrt{k}$) do little to affect these rates 6.85\% and 10.63\%, respectively. Likewise, restricting to one estimate per study has little effect 7.69\% and 10.21\%, respectively, because it can only be applied to 20 psychotherapy meta-analyses.

The not-estimable cells arise where the RE model fails to converge (a \texttt{metafor} REML non-convergence). Sladekova/A181\_5 has RE failing to converge in two of the five treatments, lowering the cell count by one (715). Including it as a change would raise the main significance rate from 7.69\% to 7.82\%; excluding it causes our main summary to be slightly more conservative.

\section{Discussion}
\label{sec:discussion}

There is a notable distinction between the size of a meta-analysis mean and how it may be broadly interpreted as statistically significant (SS) or as the smallest effect size of interest (SESOI). The mean effect size is largely robust to `outlier' handling choice; that is, the median absolute change in $d$ never exceeds 0.047 and is often an order of magnitude smaller. However, how a conclusion might be interpreted by the reader is not so robust. There are changes in SS in 7.7\% of the 715 meta-analysis-estimator combinations from at least one treatment, and 10.3\% change the SESOI status. These percentages increase further if one aggregates to meta-analyses. 11.5\% of our sample changes SS under some treatment or estimator and 15.9\% reverse their SESOI status. Recall that Langan and colleagues\cite{langan2015} found a smaller statistical significance reversal (7.5\%) among meta-analyses across alternative heterogeneity estimator choices.

In general, winsorizing has the least effect across both estimators (RE and UWLS) and treatments; DFBETAS the most, while drop-extreme and the studentized-residual treatment typically fall somewhere in between. The extremes of this ordering are expected, a priori. Winsorizing does not delete any estimate but rather shrinks 5\% of the smallest and largest estimates towards the mean. Hence, it is expected to have little effect on the size of the meta-analysis estimates of the mean effect. What little effect winsorizing has is to make a few (2.38\%) statistically nonsignificant mean estimates statistically significant by its arbitrary reduction of the heterogeneity variances. These changes run in one direction only; no result loses statistical significance (Table~\ref{tab:verdicts}). When in doubt, science prefers the more conservative conclusions, and the one-way pattern deserves attention in areas of research where statistical significance is likely already inflated by publication selection bias. In absolute counts, however, DFBETAS moves more results toward statistical significance (25) than winsorizing does (17). A deeper objection to winsorizing does not concern magnitudes at all. Winsorizing does not address the source of the problem. Estimates are pulled back to a chosen percentile without asking which ones are truly unusual, or in what sense. It may also disguise an important moderator, although this can only be judged case by case in individual meta-regressions. The case for winsorizing is equally plain. It changes the fewest conclusions of the four treatments, on both statistical significance and the SESOI assessment. A meta-analyst who wants to do something about extreme estimates without materially changing the answer can use winsorizing. We do not agree among ourselves on which consideration should prevail; this study remains agnostic by design.

In contrast, calculating and deleting notably large $|\mathrm{DFBETAS}|$ is expected to have the most effect on meta-analysis means, because this is what DFBETAS measures. That is, DFBETAS measure the difference in the meta-analysis mean estimate, calculated by either RE or UWLS, before and after that single estimate is removed from the meta-analysis sample. DFBETAS is the only one of our five registered treatments that directly assesses the robustness of the meta-analysis mean to the presence of a specific estimate in the meta-analysis sample. Thus, it is no surprise that our survey also shows that it is the most sensitive of these five `outlier' treatments. DFBETAS's sensitivity suggests that it should be the first treatment to be employed when conducting `outlier' sensitivity analysis.

Along with DFBETAS, UWLS is the most sensitive. That is, UWLS DFBETAS identifies more `outliers' and causes more changes in the potential interpretations of meta-analyses, as SS or SESOI, than RE DFBETAS. The change in the size of the mean is not notably different across these estimators (Table~\ref{tab:effectshift}). Therefore, meta-analysts who wish to further investigate specific `outliers' or their effect on meta-analysis results should calculate DFBETAS using UWLS.

\subsection{Implications for practice}

Prior research and our current survey suggest that outliers and influential estimates can influence the main findings of meta-analyses.\cite{viechtbauerCheung2010,meng2024} ``(R)esearchers generally agree that it is necessary to examine outlier and influential case diagnostics when conducting a meta-analysis'' (p.~112).\cite{viechtbauerCheung2010} Thus, it is axiomatic that these `outliers,' broadly conceived, should be identified and `handled' in some fashion. As we have seen, how meta-analysts choose to handle `outliers' can affect their conclusions; thus, it is important to preregister the exact methods and decision rules in advance of data collection.\cite{aung2026} This pre-registered study is intentionally agnostic about which `outlier' methods are best. Nonetheless, DFBETAS calculated by UWLS identified more `outliers' than RE, and they have the largest effects on the meta-analysis findings that our survey identifies. Therefore, meta-analysts who wish to consider `outlier' sensitivity need to calculate UWLS' DFBETAS. Winsorizing, at the other end, changes the fewest conclusions of the four treatments; a meta-analyst who wants a conservative check can use it. Of course, it is also defensible and consistent with our findings for meta-analysts to do more as long as their exact methods, handling rules, and robustness (or sensitivity) reporting plans are pre-registered.

But what is a meta-analyst to do when `outliers' are identified? This is, of course, the more important question. We believe that there is a consensus that identified `outliers' should not be automatically deleted as they may contain useful information needed both for the precision of the meta-analysis mean and for the identification of relevant moderators.\cite{viechtbauerCheung2010} We further agree that ``studies identified as potential outliers should always be carefully scrutinized in terms of their contents'' (p.~124).\cite{viechtbauerCheung2010} In our experience, highly influential estimates are frequently: simple errors (typos or other transcription errors), estimates of a related but nonequivalent phenomenon or measurement, interaction effects, or the effects of an important moderator variable that closer scrutiny may identify. If it is the latter, of course, the identified `outlier' should not be deleted but used in multiple meta-regression. Viechtbauer and Cheung\cite{viechtbauerCheung2010} make a similar recommendation. If, on the other hand, the `outlier' is some mistake that can be identified and documented, then this `outlier' should be corrected and retained. In the absence of an identifiable problem that can be corrected or accommodated by a moderator, the `outlier' should be omitted, and the meta-analysis results should be reported both with and without this deletion. In other words, a sensitivity analysis should be reported.\cite{viechtbauerCheung2010} Furthermore, the meta-analyst should specify in advance (pre-registration) which estimate will be the main result and which should be reported as a robustness check. Regardless, both results should be explicitly reported in the paper, and the full `outlier' identification, handling, and reporting procedure pre-registered.

In summary, a pre-registration or pre-analysis plan for `outliers' in meta-analysis should explicitly and fully specify the following steps:

\begin{enumerate}[topsep=4pt,itemsep=3pt]
\item \textit{Identifying}: Declare exactly how `outliers' are to be identified. At a minimum, $|\mathrm{DFBETAS}|$ should be used with an explicit cut-off value ($2/\sqrt{k}$) calculated by UWLS. Other methods and rules are welcome as part of a sensitivity analysis when fully and explicitly described.
\item \textit{Investigating}: State the specific ways that each identified `outlier' will be carefully examined. For example, the study from which the `outlier' has been drawn needs to be carefully re-read and the coding fully checked and verified. When a mistake can be documented, the `outlier' is corrected, and the reason reported in a verifiable way. If the `outlier' can be accommodated by a moderator, then this moderator's coding needs to be adjusted accordingly, including re-coding or checking all studies. When the `outlier' cannot be resolved in these ways, it should be omitted.
\item \textit{Reporting}: Specify which estimate of mean effect (with or without the unresolved `outlier(s)') will serve as the main finding and which is reported as the sensitivity (or robustness) analysis. Doing so further reduces researcher degrees of freedom in the ``garden of forking paths.''\cite{gelmanLoken2014}
\end{enumerate}

Some readers will, no doubt, be wondering why we did not follow our own advice in the way we handled outliers in this survey. First, the purpose of this survey is different. We are not attempting to provide the most accurate or best estimate of some mean effect, but rather to document the sensitivity of meta-analysis estimates of the mean effect to alternative, pre-registered, outlier-handling procedures across 358 behavioral science meta-analyses. Second, we pre-registered and fixed our data and our methods, handling procedures, and complete data-processing pipeline. Third, most of the BEAR~v2 collection\cite{vanzwet2026} of behavioral science meta-analyses does not contain a source/study identifier; thus, we have no way to further investigate the `outliers' as suggested in step 2, above. In our view, step 2 is often the most important step in ensuring the integrity of the resulting meta-analysis.

\subsection{Limitations}

First, the sample comes entirely from behavioral science meta-analyses, whose effect-size distributions and typical $k$ differ from those of the Cochrane Database of Systematic Reviews. Therefore, we make no claim about how these `outlier' methods and treatments will influence the typical medical, health, or other discipline's meta-analysis. Second, one random-effects meta-analysis (Sladekova/A181\_5) is not estimable under the studentized and winsorizing treatments (\texttt{metafor} REML non-convergence) and is thereby excluded. Counting it as a change in statistical significance would barely move this reversal rate (7.69\% to 7.82\%). Third, the do-nothing baseline is calculated from the public BEAR~v2 collection, which may contain meta-analyses that have already omitted some `outliers.' If an included meta-analysis had already removed or corrected unusual estimates, our baseline of `do nothing' would already carry some `outlier' handling effects. In this case, our treatment effects would understate how much these choices matter. We cannot reconstruct each source's pre-processing, but the 37 retained estimates with $|d| > 10$ (Section~\ref{sec:data}) show that at least the psychology source had not been aggressively cleaned. Fourth, 664 rows repeat a $d$ and $SE$ combination that already appears in the same meta-analysis, spread over 61 meta-analyses. In the exercise field these are distinct subsample estimates whose values coincide after rounding, but the psychology source has no study identifiers, so provenance cannot be verified there. Fifth, our meta-analyses all come from published research, where small studies tend to report imprecise and unusually large effects. The value-based treatments select estimates by magnitude, so they tend to flag these small-study estimates, and part of the change they produce may reflect small-study effects rather than the handling of true `outliers'.

\section{Conclusion}
\label{sec:conclusion}

Across 358 behavioral science meta-analyses, pre-specified outlier and influence treatments change meta-analysis mean estimates very little. However, how the meta-analysis conclusions are interpreted is sensitive to outlier and influence handling. The median absolute change in mean $d$ is at most 0.047 under any treatment or estimator, but these small shifts are enough to reverse some of the estimated means' statistical significance (SS) and smallest effect size of interest (SESOI) status. Each meta-analysis is estimated by two estimators (RE and UWLS) which gives 715 estimable meta-analysis-estimator combinations. In 7.7\% of them at least one `outlier' treatment changes SS, and 10.3\% reverse their SESOI evaluation. In other words, all five treatments agree in the remaining 92.3\% and 89.7\% interpretations across two estimators and meta-analyses. Across these 358 meta-analyses, 11.5\% reverse their SS status by some estimator and outlier treatment, while 15.9\% change their SESOI status. That is, a random draw from these behavioral science meta-analyses has a one-in-nine chance that outlier handling can change its statistical significance, and one in six for a reversal of its SESOI. Almost all of these reversals occur among results that were already close to the decision boundary. The four outlier treatments, however, have differential effects. Winsorizing reverses the fewest on both outcomes, DFBETAS the most. A reversal counts a change against the do-nothing baseline. Our study's design does not identify which treatment is correct.

Our practical advice is transparency and pre-registration. Pre-specify the outlier and influence treatments fully in the pre-registered protocol and report the central meta-analysis finding both with and without `outlier' treatment. Thus, alternative treatments are reported so that the reader can always see the sensitivity of the pre-specified outlier and influence treatment. In most cases, the difference will be practically and statistically insignificant. Where it is not, this sensitivity is itself an important meta-analysis outcome.

Our survey design and analysis pipeline can be applied to any collection of meta-analyses, and we invite others to apply them to different disciplines. Our survey and its methods inform applied meta-analysts how often the choice of outlier and influence treatments may change a meta-analysis' conclusion, offer methods researchers a framework for simulations, and give readers a simple question to ask of any borderline finding: is the central meta-analysis result robust to the handling of outliers and influential research results?

\bigskip
\noindent\textbf{Acknowledgements.} We thank the authors of the source data
this study draws together: Sladekova and colleagues for the psychology meta-analyses and
the Metapsy project for the psychotherapy meta-analyses. The exercise
meta-analyses come from Barto\v{s} and colleagues, including three of the
present authors. We are grateful to Erik van Zwet, Andrew Gelman, and Witold Wiecek for building
the harmonized BEAR collection that assembles them.

\noindent\textbf{Author contributions.}
Conceptualization: T.D.S., T.H.;
Methodology: T.H., T.D.S., Z.I., M.L.;
Software: M.L., T.H.;
Validation: T.H., Z.I., T.D.S.;
Formal analysis: M.L., T.H.;
Data curation: M.L., Z.I.;
Writing--original draft: T.H., T.D.S., M.L.;
Writing--review and editing: all authors;
Visualization: M.L., T.H.;
Supervision: T.H., T.D.S.;
Project administration: T.H., Z.I.;
Funding acquisition: T.H., Z.I., M.L.
All authors approved the final submitted draft.

\noindent\textbf{Competing interest statement.} Three of us (Havranek, Irsova,
and Luskova) are co-authors of the exercise meta-meta-analysis that supplies one
of our three source collections.\cite{bartos2025exercise} The data, the
treatments, and every threshold were fixed in the pre-analysis plan before any
result was seen. We declare no other competing interests.

\noindent\textbf{Data availability statement.} This study was pre-registered on
the Open Science Framework (\url{https://doi.org/10.17605/OSF.IO/97CMV}); the
registration holds the frozen dataset and the pre-registered analysis code, from
which the confirmatory results are push-button replicable. A complete replication
package that reproduces every table and figure in the article is archived on
Zenodo (\url{https://doi.org/10.5281/zenodo.21216506}).

\noindent\textbf{Artificial intelligence use.} Claude Opus 4.8 and Fable by
Anthropic, through Claude Code, and GPT 5.6 Sol by OpenAI, through Codex CLI
assisted in implementing the pre-registered analysis code, cross-checking the
findings, producing results tables, and editing the text. All results were
produced by running the registered code once on the frozen dataset and can be
reproduced from the public replication package. The authors are responsible for
all of the paper's content.

\clearpage

\noindent\textbf{Funding statement.} Havranek, Irsova, and Luskova acknowledge
support from the Czech Science Foundation (project 23-05227M).

\noindent\textbf{Supplementary material.} The online appendix accompanies this
article and is maintained at \url{https://meta-analysis.cz/outliers}.

\bibliographystyle{vancouver}
\bibliography{refs}

\begin{thebibliography}{10}

\bibitem{simmons2011}
Simmons JP, Nelson LD, Simonsohn U.
\newblock False-Positive Psychology: Undisclosed Flexibility in Data Collection
  and Analysis Allows Presenting Anything as Significant.
\newblock Psychological Science. 2011;22(11):1359-66.
\newblock \href {https://doi.org/10.1177/0956797611417632}
  {doi:10.1177/0956797611417632}.

\bibitem{gelmanLoken2014}
Gelman A, Loken E.
\newblock The Statistical Crisis in Science.
\newblock American Scientist. 2014;102(6):460-5.
\newblock \href {https://doi.org/10.1511/2014.111.460}
  {doi:10.1511/2014.111.460}.

\bibitem{olssonCollentine2025}
Olsson-Collentine A, van Aert RCM, Bakker M, Wicherts JM.
\newblock Meta-Analyzing the Multiverse: A Peek Under the Hood of Selective
  Reporting.
\newblock Psychological Methods. 2025;30(3):441-61.
\newblock \href {https://doi.org/10.1037/met0000559} {doi:10.1037/met0000559}.

\bibitem{steegen2016}
Steegen S, Tuerlinckx F, Gelman A, Vanpaemel W.
\newblock Increasing Transparency Through a Multiverse Analysis.
\newblock Perspectives on Psychological Science. 2016;11(5):702-12.
\newblock \href {https://doi.org/10.1177/1745691616658637}
  {doi:10.1177/1745691616658637}.

\bibitem{simonsohn2020}
Simonsohn U, Simmons JP, Nelson LD.
\newblock Specification Curve Analysis.
\newblock Nature Human Behaviour. 2020;4(11):1208-14.
\newblock \href {https://doi.org/10.1038/s41562-020-0912-z}
  {doi:10.1038/s41562-020-0912-z}.

\bibitem{patel2015}
Patel CJ, Burford B, Ioannidis JPA.
\newblock Assessment of Vibration of Effects Due to Model Specification Can
  Demonstrate the Instability of Observational Associations.
\newblock Journal of Clinical Epidemiology. 2015;68(9):1046-58.
\newblock \href {https://doi.org/10.1016/j.jclinepi.2015.05.029}
  {doi:10.1016/j.jclinepi.2015.05.029}.

\bibitem{silberzahn2018}
Silberzahn R, Uhlmann EL, Martin DP, Anselmi P, Aust F, Awtrey E, et~al.
\newblock Many Analysts, One Data Set: Making Transparent How Variations in
  Analytic Choices Affect Results.
\newblock Advances in Methods and Practices in Psychological Science.
  2018;1(3):337-56.
\newblock \href {https://doi.org/10.1177/2515245917747646}
  {doi:10.1177/2515245917747646}.

\bibitem{delgiudice2021}
Del~Giudice M, Gangestad SW.
\newblock A Traveler's Guide to the Multiverse: Promises, Pitfalls, and a
  Framework for the Evaluation of Analytic Decisions.
\newblock Advances in Methods and Practices in Psychological Science.
  2021;4(1):1-15.
\newblock \href {https://doi.org/10.1177/2515245920954925}
  {doi:10.1177/2515245920954925}.

\bibitem{viechtbauerCheung2010}
Viechtbauer W, Cheung M{\relax W-L}.
\newblock Outlier and Influence Diagnostics for Meta-Analysis.
\newblock Research Synthesis Methods. 2010;1(2):112-25.
\newblock \href {https://doi.org/10.1002/jrsm.11} {doi:10.1002/jrsm.11}.

\bibitem{aung2026}
Aung NM, Jurak I, Mehmood S, Axon E.
\newblock Sensitivity Analysis in Meta-Analysis: A Tutorial.
\newblock Cochrane Evidence Synthesis and Methods. 2026;4(1):e70067.
\newblock \href {https://doi.org/10.1002/cesm.70067} {doi:10.1002/cesm.70067}.

\bibitem{meng2024}
Meng Z, Wang J, Lin L, Wu C.
\newblock Sensitivity Analysis with Iterative Outlier Detection for Systematic
  Reviews and Meta-Analyses.
\newblock Statistics in Medicine. 2024;43(8):1549-63.
\newblock \href {https://doi.org/10.1002/sim.10008} {doi:10.1002/sim.10008}.

\bibitem{langan2015}
Langan D, Higgins JPT, Simmonds M.
\newblock An Empirical Comparison of Heterogeneity Variance Estimators in
  12\,894 Meta-Analyses.
\newblock Research Synthesis Methods. 2015;6(2):195-205.
\newblock \href {https://doi.org/10.1002/jrsm.1140} {doi:10.1002/jrsm.1140}.

\bibitem{kontopantelis2013}
Kontopantelis E, Springate DA, Reeves D.
\newblock A Re-Analysis of the {Cochrane} Library Data: The Dangers of
  Unobserved Heterogeneity in Meta-Analyses.
\newblock PLOS ONE. 2013;8(7):e69930.
\newblock \href {https://doi.org/10.1371/journal.pone.0069930}
  {doi:10.1371/journal.pone.0069930}.

\bibitem{wang2025tmeta}
Wang Y, Zhao J, Jiang F, Shi L, Pan J.
\newblock A Novel Robust Meta-Analysis Model Using the {t} Distribution for
  Outlier Accommodation and Detection.
\newblock Research Synthesis Methods. 2025;16(3):442-59.
\newblock \href {https://doi.org/10.1017/rsm.2025.8} {doi:10.1017/rsm.2025.8}.

\bibitem{noma2024}
Noma H, Sugasawa S, Furukawa TA.
\newblock Robust Inference Methods for Meta-Analysis Involving Influential
  Outlying Studies.
\newblock Statistics in Medicine. 2024;43(20):3778-91.
\newblock \href {https://doi.org/10.1002/sim.10157} {doi:10.1002/sim.10157}.

\bibitem{pap2026}
Havranek T, Irsova Z, Luskova M, Stanley TD.
\newblock Outlier and Influence Handling in Meta-Analysis: Pre-Analysis Plan;
  2026.
\newblock \url{https://doi.org/10.17605/OSF.IO/97CMV}.
\newblock Registration, Open Science Framework.

\bibitem{vanzwet2026}
van Zwet E, Gelman A, Wiecek W.
\newblock A Statistical Case for Qualified Scientific Optimism; 2026.
\newblock Working paper,
  \url{https://sites.stat.columbia.edu/gelman/research/unpublished/A_statistical_case_for_qualified_scientific_optimism.pdf}.
  Data: Benchmarks of Empirical Accuracy in Research (BEAR),
  \url{https://github.com/wwiecek/BEAR}.

\bibitem{sladekova2022}
Sladekova M, Webb LEA, Field AP.
\newblock Estimating the Change in Meta-Analytic Effect Size Estimates After
  the Application of Publication Bias Adjustment Methods.
\newblock Psychological Methods. 2023;28(3):664-86.
\newblock \href {https://doi.org/10.1037/met0000470} {doi:10.1037/met0000470}.

\bibitem{plessen2023}
Plessen CY, Karyotaki E, Miguel C, Ciharova M, Cuijpers P.
\newblock Exploring the Efficacy of Psychotherapies for Depression: A
  Multiverse Meta-Analysis.
\newblock BMJ Mental Health. 2023;26(1):e300626.
\newblock \href {https://doi.org/10.1136/bmjment-2022-300626}
  {doi:10.1136/bmjment-2022-300626}.

\bibitem{metapsy}
Harrer M, Sprenger AA, Kuper P, Karyotaki E, Cuijpers P.
\newblock {metapsyData}: Access the Meta-Analytic Psychotherapy Databases in
  {R}; 2022.
\newblock \url{https://data.metapsy.org}.
\newblock R package, Metapsy Collaboration, Vrije Universiteit Amsterdam.

\bibitem{bartos2025exercise}
Barto\v{s} F, Lu\v{s}kov\'a M, Bortnikova K, Hozov\'a K, Kantova K, Irsova Z,
  Havranek T.
\newblock Effect of Exercise on Cognition, Memory, and Executive Function: A
  Study-Level Meta-Meta-Analysis Across Populations and Exercise Categories;
  2025.
\newblock \url{https://doi.org/10.31234/osf.io/qr8e2_v1}.
\newblock Working paper.

\bibitem{viechtbauer2010}
Viechtbauer W.
\newblock Conducting Meta-Analyses in {R} with the {metafor} Package.
\newblock Journal of Statistical Software. 2010;36(3):1-48.
\newblock \href {https://doi.org/10.18637/jss.v036.i03}
  {doi:10.18637/jss.v036.i03}.

\bibitem{hartung2001}
Hartung J, Knapp G.
\newblock On Tests of the Overall Treatment Effect in Meta-Analysis with
  Normally Distributed Responses.
\newblock Statistics in Medicine. 2001;20(12):1771-82.
\newblock \href {https://doi.org/10.1002/sim.791} {doi:10.1002/sim.791}.

\bibitem{sidik2002}
Sidik K, Jonkman JN.
\newblock A Simple Confidence Interval for Meta-Analysis.
\newblock Statistics in Medicine. 2002;21(21):3153-9.
\newblock \href {https://doi.org/10.1002/sim.1262} {doi:10.1002/sim.1262}.

\bibitem{inthout2014}
IntHout J, Ioannidis JPA, Borm GF.
\newblock The Hartung-Knapp-Sidik-Jonkman Method for Random Effects
  Meta-Analysis Is Straightforward and Considerably Outperforms the Standard
  {DerSimonian}-{Laird} Method.
\newblock BMC Medical Research Methodology. 2014;14:25.
\newblock \href {https://doi.org/10.1186/1471-2288-14-25}
  {doi:10.1186/1471-2288-14-25}.

\bibitem{stanley2015}
Stanley TD, Doucouliagos H.
\newblock Neither Fixed nor Random: Weighted Least Squares Meta-Analysis.
\newblock Statistics in Medicine. 2015;34(13):2116-27.
\newblock \href {https://doi.org/10.1002/sim.6481} {doi:10.1002/sim.6481}.

\bibitem{stanley2023}
Stanley TD, Ioannidis JPA, Maier M, Doucouliagos H, Otte WM, Barto\v{s} F.
\newblock Unrestricted Weighted Least Squares Represent Medical Research Better
  than Random Effects in 67,308 {Cochrane} Meta-Analyses.
\newblock Journal of Clinical Epidemiology. 2023;157:53-8.
\newblock \href {https://doi.org/10.1016/j.jclinepi.2023.03.004}
  {doi:10.1016/j.jclinepi.2023.03.004}.

\bibitem{stanley2026uwls}
Stanley TD, Ioannidis JPA, Maier M, Doucouliagos H, Otte WM, Barto\v{s} F.
\newblock Why the Unrestricted Weighted Least Squares Should Be Routinely
  Reported in Medical Meta-Analyses; 2026.
\newblock \url{https://osf.io/vpuqj/files/4rkqg}.
\newblock Preprint.

\bibitem{stanleyHavranek2025}
Stanley TD, Doucouliagos H, Havranek T.
\newblock Reducing the Biases of the Conventional Meta-Analysis of
  Correlations.
\newblock Research Synthesis Methods. 2025;16(1):42-59.
\newblock \href {https://doi.org/10.1017/rsm.2024.5} {doi:10.1017/rsm.2024.5}.

\bibitem{cohen1988}
Cohen J.
\newblock Statistical Power Analysis for the Behavioral Sciences.
\newblock 2nd ed. Hillsdale, NJ: Lawrence Erlbaum Associates; 1988.

\bibitem{belsley1980}
Belsley DA, Kuh E, Welsch RE.
\newblock Regression Diagnostics: Identifying Influential Data and Sources of
  Collinearity.
\newblock New York: Wiley; 1980.
\newblock \href {https://doi.org/10.1002/0471725153} {doi:10.1002/0471725153}.

\bibitem{bartos2025multiverse}
Barto\v{s} F, Hoogeveen S, Sarafoglou A, Pawel S.
\newblock Single-Dataset Meta-Analysis for Many-Analysts and Multiverse
  Studies; 2025.
\newblock {arXiv}:2511.17064 [stat.ME]. \url{https://arxiv.org/abs/2511.17064}.
\newblock Preprint.

\end{thebibliography}

\end{document}